\documentclass[english,aps,prl,twocolumn,graphics,graphicx,superscriptaddress]{revtex4-1}
\usepackage[latin9]{inputenc}
\usepackage{amsmath,mathtools}
\usepackage{amssymb}
\usepackage{graphicx}
\usepackage{verbatim}
\usepackage{braket}
\usepackage{soul}
\usepackage{bm}% bold math
\usepackage{gensymb}
\usepackage{diagbox}
\usepackage{siunitx}
\usepackage[english]{babel}
\usepackage{mathtools}
\usepackage{newtxtext}  % scaled=1.005
\usepackage{amsmath}
\usepackage{amsfonts}
\usepackage{physics}
\usepackage[varbb,varg]{newtxmath} % scaled=1.005
\usepackage{bm,dsfont,eucal}

\newcommand{\B}{\mathbf}

\begin{document}
\title{Non-linear Faraday Precession of Light Polarization in Time-Reversal Invariant Materials}

\author{Falko Pientka}
\affiliation{Institut f\"ur Theoretische Physik, Goethe-Universit\"at, 60438 Frankfurt a.M., Germany}
\author{Inti Sodemann Villadiego}
\affiliation{Institut f\"ur Theoretische Physik, Universit\"at Leipzig, Br\"uderstra{\ss}e 16, 04103, Leipzig, Germany}
\date{\today} 

\begin{abstract}
We investigate the propagation of electromagnetic waves through materials displaying a non-linear Hall effect. The coupled Maxwell-Boltzmann equations for traveling waves can be mapped onto ordinary differential equations that resemble those for the motion of a pendulum. In the weakly non-linear regime relevant for most experiments, we find that the polarization of light displays a Faraday-like precession of its polarization direction that swings back and forth around the direction of Berry dipole vector as the light beam traverses the material. This occurs concomitantly with an oscillation of its degree of polarization, with a characteristic frequency that increases linearly with the intensity of the traveling wave. These effects could be observed by measuring thickness dependent Faraday rotations as well as the emission of lower frequency radiation associated with the polarization oscillations in materials displaying the non-linear Hall effect. 
\end{abstract}
\maketitle

\textcolor{blue}{\emph{Introduction.}} The Berry phases of electrons in Bloch bands can have important roles in a variety of non-linear opto-electronic phenomena \cite{belinicher1980photogalvanic,belinicher1982kinetic,von1981theory,aversa1995nonlinear,young2012first,morimoto2016topological}, that could be exploited to develop new functionalities and devices \cite{rangel2017large,cook2017design,morimoto2018current,kumar2021room,shi2023berry,onishi2024high,rappoport2023engineering,makushko2024tunable}. A phenomenon that has attracted interest among these, is the non-linear Hall effect, whereby a Hall-like current proportional to the square of the electric field can appear in metals with time-reversal symmetry controlled by the Berry curvature dipole tensor \cite{deyo2009semiclassical,moore2010confinement,sodemann2015quantum,ortix2021nonlinear,bandyopadhyay2024non}. 

However, investigations of the non-linear Hall effect have focused on viewing the electric field that drives it as being a predetermined non-dynamical variable. This letter will precisely fill in this gap by demonstrating that a beautiful and tractable set of non-linear differential equations describe the traveling electromagnetic waves in quasi-two-dimensional metals with a non-zero Berry curvature dipole. When the electric field is viewed as the coordinate of a particle, the equations resemble those of a two-dimensional pendulum subjected to an effective gravitational force pointing in a direction determined by the Berry dipole vector. As we will demonstrate, under these conditions, the non-linear Hall effect gives rise to a rocking oscillatory motion of the angle of the polarization axis, $\theta(z)$, of an elliptically polarized light beam, as it travels a distance $z$ inside the material as follows: 
\begin{align}\label{thetasimple}
 \theta(z)\approx\theta_0\cos\left(\Omega \frac{z}{\lambda}\right), \ \Omega  = \frac{\sqrt{7}}{24} \frac{I}{I_0}, \ I_0=\frac{1}{4 \alpha h}\left( \frac{D}{d}\right)^2,
\end{align}
where $I$, $\lambda$ are the light intensity and its wave-length, $I_0$ is an intensity scale determined by the Drude weight $D$ and the Berry dipole $d$ of the material, and $\theta_0$ is the initial incident angle of the polarization plane ($\alpha\approx1/137$ and $h$ are the fine structure and Planck's constants). The parameter $\Omega$ determines the angle of rotation of light polarization per traveled wavelength. We will see that the above spatial oscillations of the polarization axis are also accompanied by temporal oscillations which could be detected by the emission of low frequency radiation with a characteristic frequency given by, $\omega_{BCD}\approx \Omega \omega_{p}$, where $\omega_p$ is the plasma frequency scale of the metal. In contrast to other non-linear versions of the Faraday rotation, such as the Faraday kinetic effect (also known as current-induced optical activity) discussed in Refs.\cite{vorob1979optical,shalygin2012current,tsirkin2018gyrotropic,konig2019gyrotropic}, the above effects do not require a DC electric current to be injected or extracted into the material, and therefore can be observed in contactless optical  experiments on the materials.   

\textcolor{blue}{\emph{Electrodynamics of Non-linear Hall Plasma.}} We will focus on a 3D model of an infinite stack of 2D metallic layers with normal along ${\bf \hat{z}}$, as an approximation to layered quasi-two-dimensional materials where  large non-linear Hall effect has been observed \cite{kang2019nonlinear,ma2019observation,ma2022growth}. For electromagnetic waves traveling along ${\bf \hat{z}}$, electric currents, ${\bf j}$, and fields, ${\bf E}$, only have components along ${x,y}$, and can be described by the following set of coupled Maxwell and Boltzmann equations ($\hbar=1,e=1$):
\begin{align}
\label{Max}
    &\frac{\partial^2{\bf E}}{\partial z^2 }- \mu \varepsilon \frac{\partial^2 {\bf E}}{\partial t^2}=\frac{\mu}{w}  \frac{\partial {\bf j}}{\partial t}\\
    & {\bf j}(z,t) = -\int \frac{d^2 k}{(2 \pi )^2}f({\bf k},z,t) \frac{d \bf{r}} {d t}\\
&\partial_{t}f+\frac{d \bf{k}} {d t}\cdot\partial_{\bf k}f=0, \label{boltz}\\
& \frac{d \bf{r}} {d t}= \partial_{\bf k} \epsilon(\B k)-\frac{d \bf{k}} {d t} \times {\bf \Omega}, \ \ \frac{d \bf{k}} {d t}=- {\bf E}\label{}.
\end{align}
%\begin{align}
%\label{Max}
%    &\frac{\partial^2{\bf E}}{\partial z^2 }- \mu \epsilon \frac{\partial^2 {\bf E}}{\partial t^2}=\frac{\mu}{h}  \frac{\partial {\bf j}}{\partial t}\\
%    & {\bf j}(z,t) = -e\int \frac{d^2 k}{(2 \pi )^2}f({\bf k},z,t) \frac{d \bf{r}} {d t}\\
%&\partial_{t}f+\frac{d \bf{k}} {d t}\cdot\partial_{\bf k}f= \frac{f -f_0} {\tau}, \label{boltz}\\
%& \frac{d \bf{r}} {d t}= \partial_{\bf k} \epsilon-\frac{d \bf{k}} {d t} \times {\bf \Omega}, \ \ \frac{d \bf{k}} {d t}=-e {\bf E}\label{}.
%\end{align}
\noindent where ${\bf E}$, ${\bf j}$ only depend on $(z,t)$, $w$ is the interlayer distance, and we are taking a model of a simple single band metal with dispersion $\epsilon({\bf k})$ and Berry curvature of the form $\mathbf{\Omega}=\Omega({\bf k}) {\bf {\hat z}}$. We will be focusing on the collisionless regime relevant for plasma-like oscillations. Let us further specialize to a metal with time-reversal symmetry but no inversion symmetry, with a Berry dipole vector \cite{sodemann2015quantum} taken along the x-axis, $d=\int \frac{d^2 k}{(2 \pi )^2} f_0 \partial_{k_x} \Omega$. We specifically consider traveling waves and take the Drude weight to be diagonal within the 2D metallic planes $D_{l m}=\int \frac{d^2 k}{(2 \pi )^2} f_0\partial_{k_l}\partial_{k_m}\epsilon(\B k)=\delta_{lm} D$. By keeping terms up to second order in electric fields, and using a traveling wave ansatz, namely that fields depend only on $(v_\phi t-z)$, with $v_\phi$ the phase velocity, the above equations reduce onto a set of three ordinary differential equations analogous to those describing an effective ``particle", given by
\begin{align}
 \ddot{X}=&-X+X Y+M\dot{Y},\label{first1}\\
 \ddot{Y}=&-Y-X^2-M\dot{X},\label{second1}\\
 \dot{M}=&X,\label{third1}
\end{align}
 where $X,Y$ are dimensionless electric fields along $x,y$ directions, $M$ a dimensionless average Berry curvature, $\dot{F}=dF/d \xi$ is the derivative with respect to the dimensionless co-moving coordinate of the traveling wave, defined as
\begin{align}
 &X(\xi) \equiv E_{x}(z,t)/E_0, \
Y(\xi) \equiv E_{y}(z,t)/E_0 ,\label{Xtau}\\
& M(\xi) \equiv \frac{(v_\phi^2-c^2)^{1/2}}{w \varepsilon \omega_p v_\phi} \int \frac{d^2 k}{(2 \pi )^2}f({\bf k},z,t) \Omega({\bf k}),  \\
&\xi \equiv \frac{\omega_p (v_\phi t-z)}{(v_\phi^2-c^2)^{1/2}}, E_0 = \frac{D}{d} , \omega_p^2 = \frac{D}{w \varepsilon}, c^2 = \frac{1}{\mu\varepsilon}. \label{Mtau}
\end{align}
The above equations are obtained assuming that the phase velocity is larger than the speed of light in the material, $v_\phi>c$, which is natural in a metal, and in particular it is satisfied in the limit of usual plasma waves in the absence of the Berry curvature, which would have dispersion $\omega=(\omega_p^2+(c q)^2)^{1/2}$. It is remarkable that after performing the above rescalings Eqs.~\eqref{first1}-\eqref{third1} have no dimensionless parameter, and, therefore their behavior is entirely controlled by their dimensionless initial conditions: $\{X(0),Y(0),M(0),\dot{X}(0),\dot{Y}(0)\}$. For example, in the regime where the initial conditions are much smaller than $1$, the non-linear terms in Eq.~\eqref{first1}-\eqref{second1} can be neglected and we have the equations of a simple harmonic oscillator, which would physically correspond to the ordinary plasma waves in a metal. In the opposite, strongly non-linear regime, where the initial conditions are much larger than 1, we will see that, interestingly, Eqs.~\eqref{first1}-\eqref{third1} can be mapped exactly to those of a pendulum for certain initial conditions.

\begin{figure*}[t]
 \includegraphics[width=0.2\textwidth]{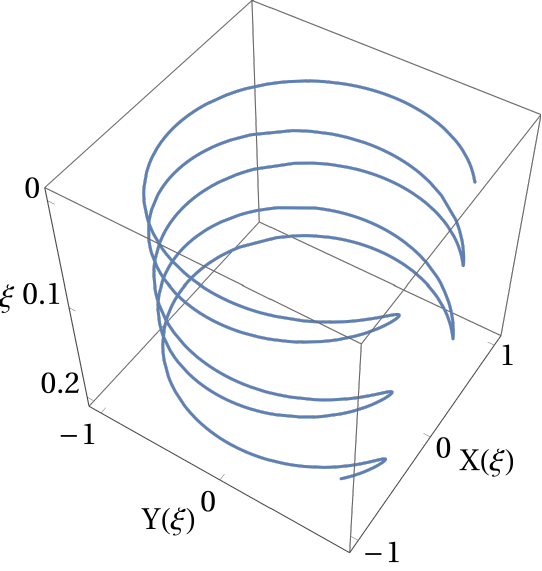}
 \includegraphics[width=0.28\textwidth]{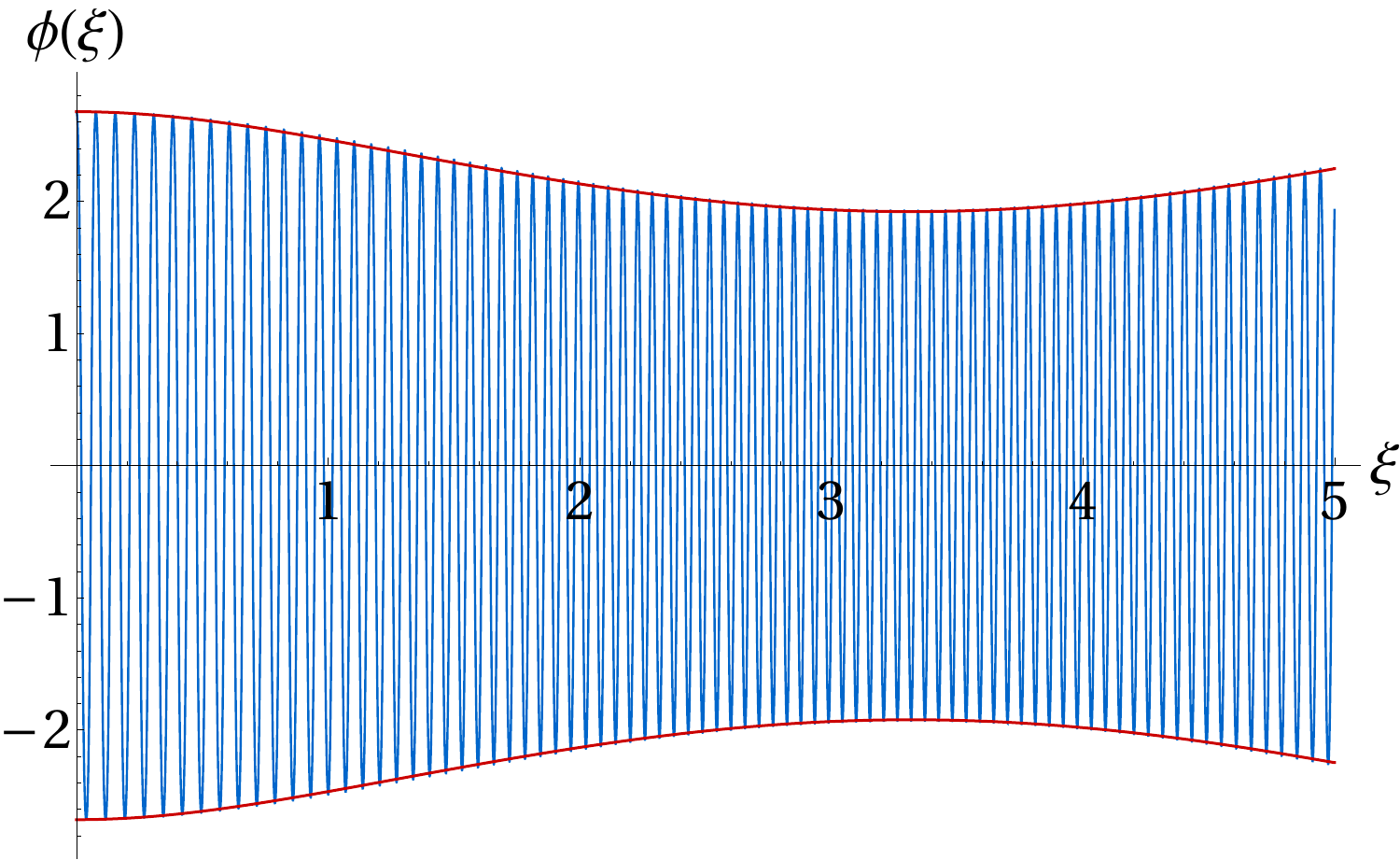}
 \includegraphics[width=0.28\textwidth]{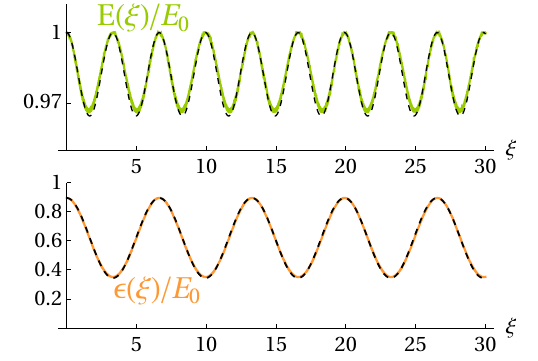}
  \includegraphics[width=0.2\textwidth]{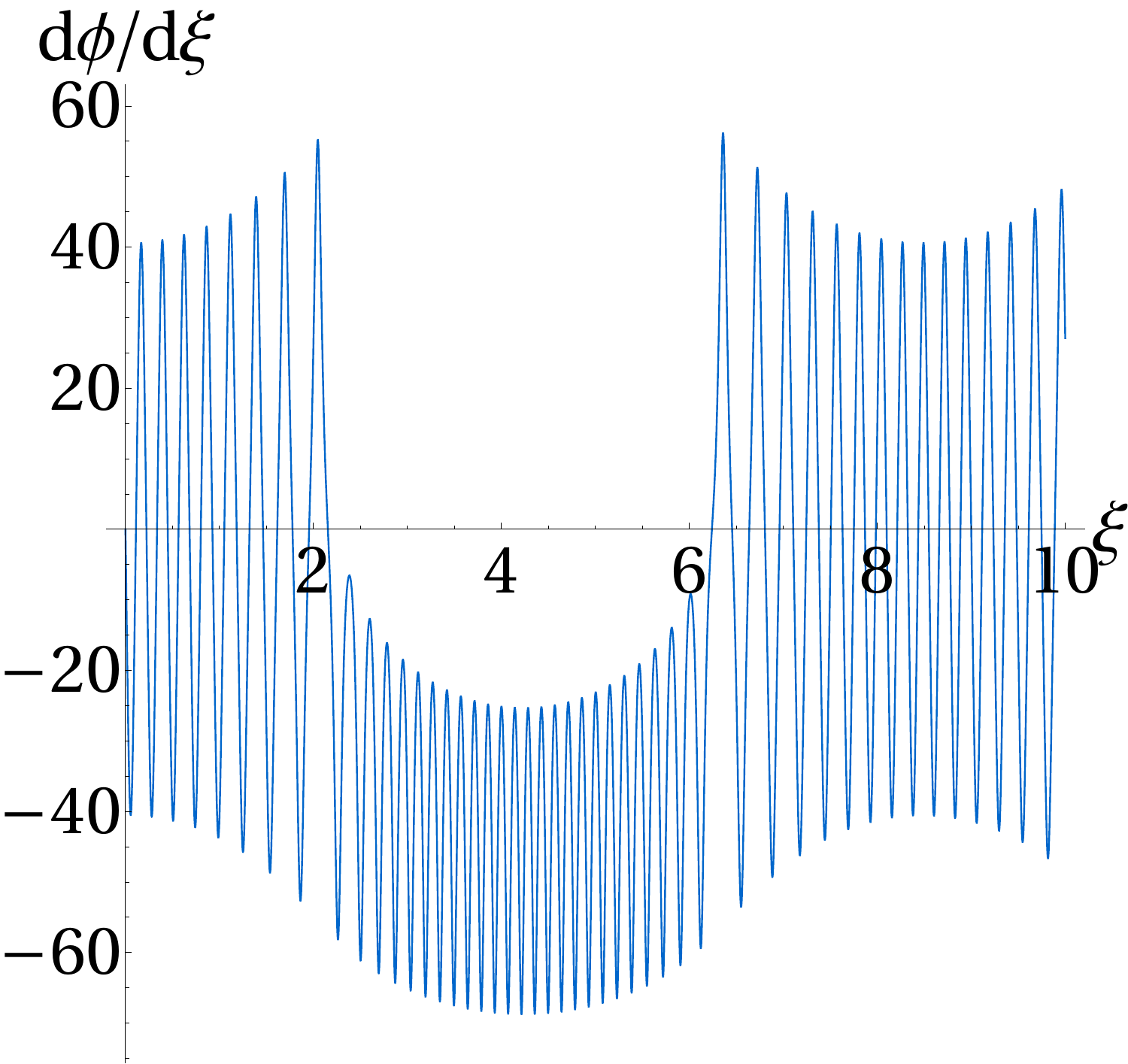}
\llap{\parbox[c]{1.0cm}{\vspace{-7cm}\hspace{-34cm}(a)}}
\llap{\parbox[c]{1.0cm}{\vspace{-7cm}\hspace{-26.5cm}(b)}}
\llap{\parbox[c]{1.0cm}{\vspace{-7cm}\hspace{-16cm}(c)}}
\llap{\parbox[c]{1.0cm}{\vspace{-7cm}\hspace{-6.5cm}(d)}}

 \caption{Numerical solution of Eqs.~(\ref{first1}-\ref{third1}) with  initial conditions in strongly non-linear regime: $Y(0)=2 \times 10^4$, $X(0)=10^4$. (a) Electric field of the travelling wave in real space.  (b) The polar angle $\phi(\xi)$ exhibtis rapid oscillations with period $\Delta \xi\simeq 0.08$. The envelope varies on much longer times scales of order one. The time dependence of the envelope is accurately captured by $\arccos(-\mathcal{E}(\xi)/E(0))$ when substituting Eq.~(\ref{sol_eps}) (red lines). (c) Electric field magnitude $E$ and energy $\mathcal{E}$ of the pendulum together with the analytical expressions in Eqs.~(\ref{sol_eps}) and (\ref{sol_E}) (black dashed lines).  The only fitting parameter is $\omega_E\simeq 0.947$, which is very close to one as expected. All other parameters are determined from the initial conditons.
 (d) Angular velocity $\dot{\phi}$ for initial conditions $y(0)=-200$, $x(0)=1000$. The pendulum is flipping in the time interval $-2<\xi<6$, where $\dot{\phi}$ has no zero crossing.
 } 
 %\caption{a}
 \label{fig:highE}
\end{figure*}

The explicit disappearance of the phase velocity $v_\phi$ in Eqs.~\eqref{first1}-\eqref{third1} also implies a kind of generalized dispersion relation between different solutions. To see this, consider a specific solution solution of Eqs.~\eqref{first1}-\eqref{third1}, with a characteristic dimensionless time scale $\xi_0$, which could be a period but it could also be some other time scale, even if the solution is not perfectly periodic. If we choose $v_\phi\rightarrow\infty$ in Eqs.~\eqref{Xtau}-\eqref{Mtau}, we would have spatially uniform fields with zero wave-vector and characteristic frequency scale $\omega_0= 2\pi\omega_p/\xi_0$. The time scales of these uniform solutions can be viewed as modified non-linear plasma scales. By choosing a finite $v_\phi$ in Eqs.~\eqref{Xtau}-\eqref{Mtau}, instead, we can construct a new solution with finite wave-vector from the original spatially uniform solution. In general, its time scales will depend on the characteristic wave-vector, $q$, and frequency $\omega_q=v_\phi q$, which are related to the frequency of the $q=0$ solution as follows:
\begin{align}
 \omega_q=\sqrt{\omega_0^2+(c q)^2},
\end{align}
The key difference with respect to the linear case, however, is that there is no unique non-linear frequency scale, $\omega_0$, associated with the solutions of Eqs.~\eqref{first1}-\eqref{third1}. In general its time scales will depend on initial conditions, analogously to how the period of a pendulum depends on its initial energy.

In the limit of large amplitudes, when the linear terms in Eqs.~\eqref{first1}-\eqref{third1} can be neglected, there is an additional transformation that relates different solutions of Eqs.~\eqref{first1}-\eqref{third1} with different characteristic times, $\xi_0$, and amplitudes. Namely, given a solution $(X(\xi),Y(\xi),M(\xi))$, the following re-scaling, leads to a new solution: $(X'(\xi),Y'(\xi),M'(\xi))=(\kappa X(\sqrt{\kappa}\xi),\kappa Y(\sqrt{\kappa}\xi),\sqrt{\kappa} M(\sqrt{\kappa} \xi))$. Therefore, this generates a new set of solutions with rescaled dimensionless time $\xi_\kappa=\xi_0/\sqrt{\kappa}$, from a given solution with time scale $\xi_0$.

%\begin{align}
%\left(
%\begin{array}{c}
% M(\xi)  \\
%X(\xi) \\
% Y(\xi) \\
%\end{array}
%\right)\to \left(
%\begin{array}{c}
% \sqrt{\kappa} \ M(\sqrt{\kappa} \xi) \\
% \kappa \ X(\sqrt{\kappa}  \xi)\\
%\kappa \ Y(\sqrt{\kappa}  \xi) \\
%\end{array}
%\right),
%\end{align}

%\section{Traveling-wave solution}

\textcolor{blue}{\emph{Traveling waves in strongly non-linear regime.}} In the following, we solve Eqs.~(\ref{first1})-(\ref{third1}) for the case that the initial conditions do not break time-reversal symmetry, $\dot{X}(0)=\dot{Y}(0)=M(0)=0$, relevant for time-reversal invariant systems.  In order to illustrate more clearly the impact of the Berry curvature dipole on the dynamics, we first set the Drude weight to zero,
\begin{align}
 \ddot{X}=&XY+M\dot{Y},\label{first}\\
 \ddot{Y}=&-X^2-M\dot{X},\label{second}\\
 \dot{M}=&X.\label{third}
\end{align}
We will denote the dimensionless electric field vector by $\B E=(X,Y)$, and its magnitude by $E=\sqrt{X^2+Y^2}$. Equations~(\ref{second}) and (\ref{third}) lead to $d(MX)/d\xi=-\ddot{Y}$, which yields  $MX=-\dot{Y}$ after integration. Similarly we obtain $MY=\dot{X}$ from Eqs.~(\ref{first}) and (\ref{third}).  These two relations lead to $\B E\cdot \dot{\B E}=X\dot{X}+Y\dot{Y}=0$, implying that the electric field amplitude, $E$, is time-independent. In polar coordinates,  $Y=-E \cos\phi$, $ X=E \sin\phi$,
Eq.~(\ref{second}) can be written as
\begin{align}
 \ddot{\phi}+E \sin\phi=0,
\end{align}
which is the equation of motion for a pendulum with a length $\sim 1/E$. Hence, if the electric field was only affected by the terms originating from the Berry dipole, it would follow a pendulum motion with gravity along the $-y$ direction. The pendulum motion is illustrated by a real space representation of the traveling wave in Fig.~\ref{fig:highE}(a).
The pendulum solution turns out to be unphysical, as it allows for an average electric field $E_y$. This is a consequence of ignoring the restoring force represented by the Drude weight. We can account for the restoring force by considering Eqs.~(\ref{first1})-(\ref{third1}) in the strongly nonlinear limit, $E\gg1$. In this limit the Berry dipole terms dominate and the electric field follows a pendulum motion on short time scales $\xi\sim O(1/\sqrt{E})$. The dynamics on longer time scales can be understood by considering the time average, $\langle Y \rangle$, over a pendulum period $T$ and using Eq.~(\ref{second1}), we obtain
\begin{align}
\langle \ddot{Y} \rangle+ \langle Y \rangle\simeq\int_{T} d\xi (-X+\dot{M})X=0,
\end{align}
where we have used partial integration and Eq.~(\ref{third1}). Hence the average height of the pendulum evolves like a harmonic oscillator, and we have
$ \langle Y \rangle \simeq Y_0\cos(\omega_\xi \xi)$, where $\omega_\xi\simeq 1$ and $Y_{0}$ is a constant. Hence even a small Drude weight is enough to ensure that the electric field averages to zero over the  Plasma timescale. In the language of the pendulum, the linear terms in Eqs.~(\ref{first1}) and (\ref{second1}) lead to a periodic modulation of the pendulum  energy ${\cal E}=\dot{\phi}^2/2+Y$. An analytical treatment of the strongly nonlinear case (see Supp. Sect. \ref{strong-supp}) shows that the length of the pendulum is also modulated periodically with twice the period. The numerical solution is shown along with the analytical estimate in Fig.~\ref{fig:highE}(b) and (c).
Interestingly, the vanishing of the average field implies that in some cases the pendulum may  flip for a finite number of consecutive cycles, namely when the pendulum is initially sufficiently close to the minimum (see Fig.~\ref{fig:highE}(d)). Such solutions are nevertheless time-reversal invariant as the rotation direction alternates. 

%CONTINUE

%We now examine how this dynamics is changed by the Drude weight term
%\begin{align}
% \ddot{x}=&-x+xy+m\dot{y}\label{first1}\\
% \ddot{y}=&-y-x^2-m\dot{x}\label{second1}\\
% \dot{m}=&x\label{third1}
%\end{align}
%with the same initial conditions. From Eqs.~(\ref{first1}) and (\ref{third1}) we find  $d[m(y-1)]/dt=\ddot{x}$ and thus
%\begin{align}
% m= \frac{\dot{x}}{y-1}\label{msol}
%\end{align}

\begin{figure*}[t]
 \includegraphics[width=0.22\textwidth]{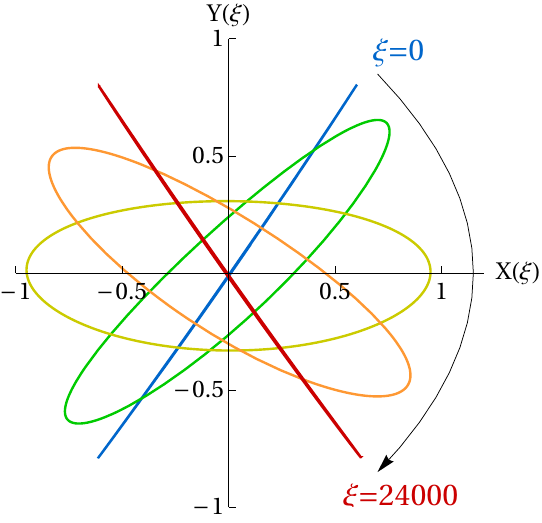}
 \includegraphics[width=0.37\textwidth]{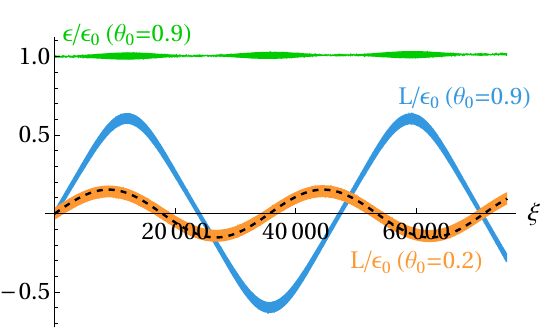}
 \hspace{.01\textwidth}
 \includegraphics[width=0.35\textwidth]{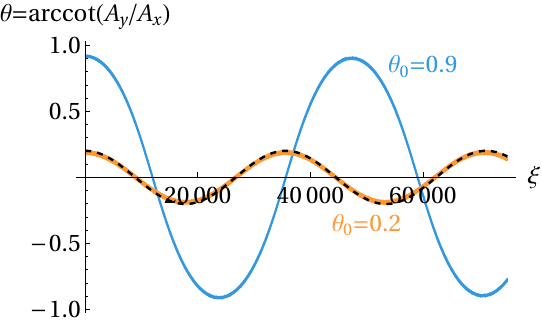}
\llap{\parbox[c]{1.0cm}{\vspace{-7.2cm}\hspace{-34cm}(a)}}
\llap{\parbox[c]{1.0cm}{\vspace{-7.2cm}\hspace{-25.5cm}(b)}}
\llap{\parbox[c]{1.0cm}{\vspace{-7.2cm}\hspace{-12.7cm}(c)}}

 \caption{Numerical solution of Eqs.~(\ref{first1}-\ref{third1}) with initial conditions in weakly non-linear regime: $y(0)=a_0 \cos\theta_0$, $x(0)=a_0 \sin\theta_0$,  $a_0=0.04$. (a) Electric field during one period $\xi \to \xi+2\pi$ at different points in time. Evolution of the  semi-major axis is indicated by the arrow. The times are $\xi=\{0,6,12,16,24\}\times 10^3$ from blue to red along the arrow. At longer times, the trajectory is traced backwards until initial direction is recovered at $\xi\simeq 48000$. (b) Effective energy and angular momentum normalized by initial energy $\epsilon_0=a_0^2/2$. (c) Angle of semi-major axis determined from Laplace-Runge-Lenz vector. The energy is almost constant in agreement with our analytical results. The black dashed lines in (b) and (c) indicate the analytical result for small initial angles given by Eqs.~(\ref{Lsol1}) and (\ref{theta_sol}). The apparent thickness of curves in (b) and (c) arises because of short time oscillations, whose amplitude decreases as $E\to0$ (the short time scale $\delta \xi=2\pi$ is not resolved on the scale of the plots).
 } 
% \caption{a}
 \label{fig:smallE}
\end{figure*}

\textcolor{blue}{\emph{Traveling waves in weakly non-linear regime.}} For typical materials the light intensity scale, $I_0$, that separates the weakly and strongly non-linear regimes is very large (i.e. $I_0 \sim 10^{13} W/m^2$, see discussion section), and thus it is more realistic to consider the opposite regime of weak non-linearity, where it is convenient to express Eqs.~\eqref{first1}-\eqref{third1} in vector notation,
\begin{align}
\ddot{\B E }+\B E = -X \hat{z}\times \B E+\frac{\dot{X}}{1-Y} \hat{z}\times \dot{\B E},\label{eom}
\end{align}
Here $M$ has been eliminated using $M=\dot{X}/(Y-1)$, which follows from substituting Eq.~\eqref{third1} into Eq.~\eqref{first1} and then integrating. We now have a different separation of time scales, where the left-hand side, which dominates on short time scales, describes the evolution of a harmonic oscillator associated with standard plasma oscillations. In these short time scales the electric field vector traces an ellipse. On longer time scales, the an-harmonic terms on the right-hand side originating from the Berry curvature dipole become important. In order to understand the long-time evolution, we consider suitable constants of motion of the 2d harmonic oscillator, which can only change due to the nonlinear terms. We consider its energy $\epsilon$, angular momentum ${ L}$ and the direction of the Laplace-Runge-Lenz vector $\B A$ (see Ref. Fradkin 1966),
\begin{align}\label{conserved}
\epsilon&=\frac{ E^2+(\dot{\B E})^2}{2},\qquad  L= E^2\dot{\phi},\\
 \B A&=s\frac{\dot{\B E}\times \hat{z}  L+\B E(\sqrt{\epsilon^2-L^2}-\epsilon)}{\sqrt{E^2(\epsilon^2-{ L}^2)^{1/2}-E^2\epsilon^2+{ L}^2}}\label{Runge},
\end{align}
where $s=\pm1$ is chosen such that $A_y>0$.
For an elliptic trajectory with semi-major and semi-minor axis $a$ and $b$, we have $\epsilon=(a^2+b^2)/2$, $| L|=ab$, and $\B A$ points along the semi-minor axis. Linear polarization corresponds to $ L=0$ and circular to $| L|=\epsilon$. The sign of $ L$ indicates the direction of the polarization rotation.

Figure~\ref{fig:smallE}(a) illustrates one of our central findings, namely, it shows the trajectories traced by the electric field during one cycle, $\Delta \xi=2\pi$, of the harmonic oscillator (i.e. the plasma time-scale) at different points during the long-time evolution. The light is initially linearly polarized then changes to elliptic polarization and back to linear polarization. The direction of the semi-major axis rotates in the plane until it reaches the mirror image of the initial direction  with respect to one of the x-axis, where the light is again linearly polarized. Subsequently, this path is traced backwards until the initial condition is reached again. This rocking motion of the polarization can be viewed as taking place either in time at some fixed plane, or as a function of position traveled by the wave, and can be used to convert linearly polarized light into elliptically polarized, or rotate its axis of polarization.

The energy of the harmonic oscillator, which corresponds to the intensity of light, is a constant even on long time scales as shown in Fig.~\ref{fig:smallE}(b) (ignoring the weak fluctuations we find no appreciable change of $\epsilon$). This is a reflection of the fact that the Berry curvature dipole contribution to the response is nondissipative. The change between linear and elliptic polarization is captured by the angular momentum also shown on Fig.~\ref{fig:smallE}(b), which oscillates around zero on time scales $\xi \sim 1/E^2 $. We see that the elliptic polarization has opposite rotation direction of the electric field vector during the first and second half of the period. Moreover we have that $| L|<\epsilon$ even at the maximum, meaning that circular polarization is never reached.
The polarization direction, i.e., the angle of the semi-major axis $\theta={\rm arccot}\, A_y/A_x$, is shown in Fig.~\ref{fig:smallE}(c). The polarization direction swings around the x-axis between the initial point and it's mirror image, somewhat similarly to the pendulum motion of the electric field discussed in the strongly nonlinear case, however now the direction of the Berry dipole acts as the direction of the effective gravity. 

We can derive an approximate analytical solution to Eq.~\ref{eom} in the limit of small amplitudes, $E\ll 1$, and small initial deviations of the polarization direction, $\theta_0\ll1$, from the x-axis around which it performs its rocking motion. To lowest order in the electric field the solution reads 
\begin{align}
X_0+iY_0&=(a_0 \cos \xi+ib(\xi) \sin \xi) e^{i\theta(\xi)}\label{sol2a}
\end{align}
with $X_0(0)+iY_0(0)=a_0 e^{i \theta_0}$. The polarization vector describes an ellipse, whose semi-major axis direction has effective time-dependent polar angle $\theta(\xi)$, and constant length, $a_0$, while the length of its semi-minor axis, $b(\xi)$, changes in time. Notice that the elliptic motion is traced on short time scales of order $\xi\sim 1$, whereas $\theta(\xi)$ and $b(\xi)$ vary on much longer timescales $\xi\sim 1/E^2$. To obtain the long-time dynamics of these quantities, we expand the general solution to second order in $E$ as $X+i Y \approx (1+\frac{i}{3}(X_0-\dot{X}_0))(X_0+iY_0)$. We can use this expression to calculate the time derivative of the constants of motion of the 2d harmonic oscillator, like the angular momentum or the energy of the motion along $y$, $\epsilon_y=(Y^2+\dot{Y}^2)/2$, (see Suppl. Sec. \ref{weak-supp}), and we obtain
\begin{align}
    b(\xi)=\frac{a_0\theta_0}{\sqrt{7}}\sin(\Omega \xi),\theta(\xi)=\theta_0\cos(\Omega \xi), \Omega=\frac{a_0^2\sqrt{7}}{24}.\label{theta_sol}
\end{align}
%Here $\xi$ is the effective dimensionless time-like variable defined in Eq.~\eqref{Mtau}. 
To leading order $a_0=E(0)$ is the dimensionless magnitude of the electric field, which we used to derive the simplified approximate expressions presented in Eq.~\eqref{thetasimple}. As illustrated in Fig.~\ref{fig:smallE}, the observables calculated from these analytical expressions are in excellent quantitative agreement with the numerical results at small angles and amplitudes. For larger initial angles, there is still qualitative agreement, although the period is somewhat longer and the angular momentum and the polar angle deviate from a single-harmonic time dependence.

\textcolor{blue}{\emph{Discussion and outlook for experimental detection.}} We have demonstrated that when light travels in the direction perpendicular to a stack of 2D metals with a non-linear Hall effect, its plane of polarization oscillates back and forth like a pendulum. There is a variety of van-der-Waals layered materials where strong non-linear Hall effects has been detected, such as WTe$_2$ \cite{kang2019nonlinear,ma2019observation}, MoTe$_2$ \cite{ma2022growth} and TaIrTe$_4$ \cite{kumar2021room} which are prime candidates to observe this non-linear Faraday precession. As shown in Eq.~\eqref{thetasimple}, the polarization rotation per traveled wave-length is controlled by the dimensionless parameter $\Omega$, which can be equivalently expressed in terms of electric fields as $\Omega  = (\sqrt{7}/24) (E/E_0)^2$. Here $E_0=\sigma/\sigma_{\rm NLH}$ is an material dependent electric field scale determined by the ratio of the linear Drude conductivity,  $\sigma$, and the non-linear Hall conductivity, $\sigma_{\rm NLH}$. This  parameter can be easily measured in a Hall bar geometry \cite{kang2019nonlinear,ma2022growth}. For example, measurements in a MoTe$_2$ bilayer \cite{ma2022growth} yield a scale $E_0\approx 10^8 V/m$, which corresponds to a light intensity $I_0= E_0^2 (e^2/(4 \alpha h))\approx 3\times 10^{13} W/m^2$. Thus, for radiation with intensity $10^{8} W/m^2$ one gets a rotation rate of the plane of polarization of about $\Omega/2\pi \sim 4\times10^{-7}$ radians per traveled wavelength, which increases linearly with the intensity of radiation. Therefore this non-linear Faraday precession of polarization appears to be  comfortably within the detection limits of modern interferometers, which can have resolutions of $10^{-9}$ radians (see e.g. \cite{kapitulnik2009polar}). The rocking of the polarization plane as a function of distance could be detected by measuring samples of different thicknesses. These experimental studies, could therefore uncover this novel type of non-linear Faraday precession present in time-reversal invariant metals, which would offer a new purely optical technique to detect their breaking of inversion symmetry and probe the Berry curvature distribution around their Fermi surface.

\textit{Acknowledgements}--
We thank Ivo Souza for useful correspondence. We acknowledge support from the Deutsche Forschungsgemeinschaft (DFG) through research grant project number 555335098.

%\bibliography{references}

\clearpage
\widetext

\begin{center}
\textbf{\large Non-linear Faraday Precession of Light Polarization in Time-Reversal Invariant Materials}
\end{center}

\setcounter{equation}{0}
\setcounter{figure}{0}
\setcounter{table}{0}
\setcounter{page}{1}
\makeatletter
\renewcommand{\theequation}{S\arabic{equation}}
\renewcommand{\thefigure}{S\arabic{figure}}

\newcounter{proplabel}

\section{Traveling waves in strongly non-linear regime}\label{strong-supp}

Let us first consider large electric field amplitudes $E\gg 1$. Using dimensional analysis we can understand the dynamics of the electric field vector as a combination of two processes happening on different timescales. On short times scales of order $E^{-1/2}$, the linear Drude weight terms can be ignored and the electric field vector follows a pendulum motion with $\dot{\phi}\sim O(E^{1/2})$ and $E\approx {\rm const.}$ as discussed in the main text.  This is evidenced by the time evolution of the polar angle shown in Fig.~\ref{fig:highE}(b), which exhibits fast oscillation on the scale $\Delta \xi\sim E^{-1/2}$. 
However, the linear terms cause an evolution of the properties of the pendulum on longer time scales of order one (i.e. the plasma time-scale) as can be seen from the variation of the turning points of the pendulum (i.e. the envelope of the oscillation). The long time dynamics can be captured by considering the energy of the pendulum
\begin{align}
 \mathcal{E}(\xi)=\frac{1}{2}\dot{\phi}^2+Y\label{energy},
\end{align}
and the electric field strength $E$ corresponding to its inverse length which are both conserved on short time scales. The plot of their time evolution in Fig.~\ref{fig:highE}(c) exhibits oscillations with period $\sim 2\pi$ and $\sim \pi$, respectively. Hence the long time dynamics of the electric field vector appears to be periodic as well. As it turns out, such a simple pendulum motion with oscillating turning point only takes place for a sufficiently large initial angle $\phi$. If the initial energy is small, in a sense to be made more precise below, the pendulum can gain enough energy over time to flip meaning the electric field vector rotates. This can be seen in a plot of the angular velocity in Fig.~\ref{fig:highE}(d) for a smaller initial amplitude, where an initial pendulum motion is followed by a rotation evidenced by a negative angular velocity over an extended period of time.

%\begin{figure*}[t]
 %\includegraphics[width=0.2\textwidth]{xandy}
 %\includegraphics[width=0.28\textwidth]{phit}
 %\includegraphics[width=0.28\textwidth]{epst}
  %\includegraphics[width=0.2\textwidth]{dphidt}

 %\caption{Numerical solution of Eqs.~(\ref{first1}-\ref{third1}) with initial conditions for (a-c): $y(0)=2 \times 10^4$, $x(0)=10^4$. (a) Electric field vector of the travelling wave solution in real space.  (b) The polar angle $\phi(t)$ exhibtis rapid oscillations with a period $\Delta t\simeq 0.08$. The envelope varies on much longer times scales of order one. The time dependence of the envelope is accurately captured by $\arccos(-\mathcal{E}(t)/E(0))$ when substituting Eq.~(\ref{sol_eps}) (red lines) . (c) Electric field strength $E$ and energy $\mathcal{E}$ of the pendulum together with the analytical expressions in Eqs.~(\ref{sol_eps}) and (\ref{sol_E}) (black dashed lines).  The only fitting parameter is $\omega_E\simeq 0.947$, which is very close to one as expected. All other parameters are determined from the initial conditons.
 %(d) Angular velocity $\dot{\phi}$ for initial conditions $y(0)=-200$, $x(0)=1000$. The pendulum is flipping in the time interval $-2<t<6$, where $\dot{\phi}$ has no zero crossing.
 %} 
% \caption{a}
 %\label{fig:highE}
%\end{figure*}

In the following, we focus on sufficiently large initial angles $\phi$ such that full swings of the pendulum never occur and a turning point with $\dot{\phi}=0$ exists during every short-time interval of order $E^{-1/2}$. 
To make analytical progress, we express the equations of motion in polar coordinates
\begin{align}
  \ddot{E}+E=&E\dot{\phi}M+E\dot{\phi}^2.\label{E_harmonic_oscillator}\\
  \ddot{\phi}+X=&-\frac{\dot{E}(2\dot{\phi}+M)}{E}.\label{dtdm}
\end{align}
Replacing $M$ using $M=\dot{X}/(Y-1)$, we can rewrite Eq.~(\ref{E_harmonic_oscillator}) in the limit $Y\gg1$ after some algebra as
\begin{align}
\frac{Y^2}{E^2}\frac{d}{d\xi}\Bigl(\frac{\dot{E}}{Y}\Bigr) \simeq&- \frac{\dot{\phi}^2+Y}{E}-\frac{\dot{E}^2Y}{E^3}.
\end{align}
Because of $Y\sim \dot{\phi}^2\sim O(E)$, dimensional analysis implies that  $\dot{E}\sim O(E)$ and $d(\dot{E}/Y)/d\xi\sim O(1)$. This means the short-time behavior of $\dot{E}$ is entirely determined by the temporal dependence of $Y(\xi)$ and hence we can write 
\begin{align}
 \dot{E}(\xi)=f_{\rm slow}(\xi)Y(\xi)+f_{\rm fast}(\xi)\label{dEdt}
\end{align}
where $f_{\rm slow}\sim O(1)$ is a function that varies slowly $\dot{f}_{\rm slow}\sim O(1)$ and  $f_{\rm fast}$ is a small correction to the first term varying on short timescales such that $\dot{f}_{\rm fast} \sim \dot{\phi}f_{\rm fast}$ is of order $E$ or less. 
From $\dot{E}(\xi=0)=0$ we deduce $f_{\rm slow}(\xi=0)\simeq0$. We are now ready to evaluate the time evolution of the pendulum energy to leading order in $1/E$,
\begin{align}
 \dot{ \mathcal{E}}\simeq\frac{f_{\rm slow}(Y^2-Y\dot{\phi}^2)}{E},
\end{align}
where we have used $M=\dot{X}/(Y-1)$ and Eqs.~(\ref{dtdm}) and (\ref{dEdt}). 

As argued above we expect the energy and length of the pendulum to be approximately conserved on short time scales. We therefore proceed with a Born--Oppenheimer type approximation, considering time averages over a short cycle between two turning points with period $\Delta \xi\sim E^{-1/2}$, i.e.,
\begin{align}
 \dot{ \mathcal{E}}&\simeq\frac{f_{\rm slow}}{E\Delta \xi}\int_{\Delta\xi} d\xi (Y^2-Y\dot{\phi}^2),
\end{align}
where we have assumed that $E$ is roughly constant within one period.
Recognizing that $\dot{X}\simeq -Y\dot{\phi}$ to leading order and doing a partial integration, we find
\begin{align}
 \dot{ \mathcal{E}}\simeq\frac{f_{\rm slow}}{E\Delta t}\int_{\Delta \xi} d\xi (Y^2-X\ddot{\phi}).
\end{align}
Finally, we can substitute the equation of motion $\ddot{\phi}\simeq -X$ as well as $E^2=Y^2+X^2$ and we arrive at 
\begin{align}
 \dot{ \mathcal{E}}
 &\simeq f_{\rm slow}E,
\end{align}
with $  \dot{ \mathcal{E}} (0)\simeq 0$ . The field strength $E$ can be obtained from Eq.~(\ref{dEdt}) to leading order as
\begin{align}
 E(\xi)\simeq E_0+\int_0^\xi d\xi'f_{\rm slow}(\xi')Y(\xi').
\end{align}
where, as we will see later, the constant first term always dominates the oscillatory second term. We plug this into the previous equation, keeping only the first term, and obtain a slowly varying energy
\begin{align}
\mathcal{E}(\xi) &\simeq \mathcal{E}_0+E_0\int^\xi d\xi'f_{\rm slow}(\xi')\label{eps_result}.
\end{align}
A pendulum's energy is related to its average height $\langle Y\rangle$ by a monotonously increasing function, which we will linearize below (e.g., for small initial angles $|\phi|\ll1$ we have $\mathcal{E}\simeq2 \langle Y\rangle+E$). The average height, which has been determined in the main text,
\begin{comment}
can be determined by taking the average over one pendulum period of Eq.~(\ref{second1})
\begin{align}
 \langle \ddot{Y}\rangle+\langle Y\rangle=\int_{\Delta \xi} d\xi(-X^2-M\dot{X})\simeq\int_{\Delta t} d\xi(-X+\dot{M})X=0,
\end{align}
where we have used partial integration and Eq.~(\ref{third1}). Hence the average height 
\end{comment}
evolves like a harmonic oscillator, 
$ \langle Y\rangle\simeq Y_0\cos(\omega_E \xi)$, where $\omega_E\simeq 1$ and $Y_{0}$ is a constant. Here we have used the initial condition $(d\langle Y\rangle/d\xi)(\xi=0)\simeq 0$, which follows from $\dot{\mathcal{E}}(\xi=0)\simeq 0$.
Small deviations of the frequency from one can occur because we have ignored the boundary term resulting from partial integration, which is only approximately zero. Assuming a linearized dependence of $\mathcal{E}$ on $\langle Y\rangle$ (which is a reasonable approximation except for initial angles close to the maximum) and using Eq.~(\ref{eps_result}) we find  
\begin{align}
f_{\rm slow}(\xi)\simeq -f_0 \sin (\omega_E \xi), 
\end{align}
with $f_0$ independent of time.
We thus obtain for the energy and field strength 
\begin{align}
 \mathcal{E}(\xi) &\simeq  \mathcal{E}_0+E_0 f_0 \cos(\omega_E \xi)\label{sol_eps},\\
 E(\xi)&\simeq E_0+\frac{f_0 Y_0}{4} \cos(2 \omega_E \xi).\label{sol_E}
\end{align}
The parameters $Y_0$, $f_0$, $ \mathcal{E}_0$ and $E_0$ can be determined directly from the initial length and energy of the pendulum without solving the full time dependence. For instance, the parameter $f_0$ can be determined from the difference between the initial energy of the pendulum and the energy of a pendulum with average height $\langle Y\rangle=-Y_0$. Because the derivative of $ \mathcal{E}$ and $\langle Y\rangle$ must have the same sign, $f_0$ and $Y_0$ have the same sign. In particular this implies that the field strength $E$ is always smaller or equal to the initial value. Because $|Y_0|< E$ and $f_0\sim Y_0$ we see that $E_0$ typically dominates over the oscillatory term in Eq.~(\ref{sol_E}).

In summary, the electric field follows the motion of a pendulum on short time scales $\sim 1/\sqrt{E}$ with gravity along the $-y$ direction. On longer times scales, there are two effects: the angle of the turning points changes periodically with a frequency $\omega_E\simeq 1$ and the field amplitude (i.e., the inverse length of the pendulum) varies by a small amount with frequency $2\omega_E$. Fig.~\ref{fig:highE} demonstrates that this solution quite accurately describes the numerics in the limit of large field strengths, $E\gg 1$, and provided that the pendulum never flips. Flips cannot occur for sufficiently large initial angles, when the parameter $Y_0$ (the average pendulum's height during the first few cycles) is positive. In this case $f_0$ is also positive and hence the energy $ \mathcal{E}$ (see Eq.~\ref{sol_eps}) is always smaller than its initial value and the pendulum cannot flip. On the contrary if $Y_0$ is negative initially, its energy increases over time and flips cannot be excluded (and typically occur). Hence initial values with $Y_0>0$ roughly delineate the regime of validity of our analytical solution.

\section{Traveling waves in weakly non-linear regime}\label{weak-supp}

We now consider the opposite limit of very weak electric fields, where the nonlinear terms in the right hand side of Eq.~\eqref{eom} can be considered a perturbation. We will analytically estimate  the time evolution of the oscillator constants of motion, described in Eqs.~\eqref{conserved} and \eqref{Runge}, in the limit $E\ll 1$. To eliminate the effect of fast fluctuations with a small amplitude, we consider the short time averages    $\langle f(\xi) \rangle=\int_\xi^{\xi+2\pi} d\xi' f(\xi')/2\pi$. Since quantities like $L$ and $\epsilon$ are second order in $E$ and we are interested in the dynamics on time scales $\xi \sim E^{-2}$, we need to expand their time derivatives at least to fourth order in $E$. We start by solving the equation of motion (\ref{eom}) order-by-order in the small parameter $\B E=(X,Y)$. Up to second order we obtain 
\begin{align}
X+iY&=\Bigl[1+\frac{i}{3}(X_0-\dot{X}_0)\Bigr](X_0+iY_0),\label{sol2}\\
X_0+iY_0&=(a \cos \xi+ib \sin \xi) e^{i\theta},\label{sol2a}
\end{align}
where the second line defines an ellipsis, whose semi-major axis has a polar angle $\theta$. The intial condition are $a(0)=a_0$, $b(0)=0$, and $\theta(0)=\theta_0$. In this parametrization, we obtain to leading order
\begin{align}
    \epsilon=(a^2+b^2)/2,\qquad L=ab
\end{align}
This solution with constant parameters $a,b\sim O(E)$ and $\theta$ is valid only on short time scales $\xi \sim 1$. However, we can use it to determine the long-time evolution of the oscillator constants of motion. For the energy we obtain $\dot\epsilon=\dot{\B E}(\ddot{\B E}+\B E)=-XL$ from Eq.~(\ref{eom}) and, using the expressions in Eqs.~(\ref{sol2}) and (\ref{sol2a}), we have $\langle \dot\epsilon\rangle=0$ at least to order $E^4$, which means the energy remains indeed constant (except for small fluctuations) on long time scales of order $\xi \sim1/E^2$. For the angular momentum, Eq.~(\ref{eom}) yields to fourth order
\begin{align}
    \dot{ L}&=\hat{z}\cdot (\B E\times \ddot{\B E})=-X E^2+(1+Y)\dot{X}\dot{E} E.\label{L_dot}\\
    \langle \dot{ L}\rangle&=\frac{1}{48}(a^4-b^4)\sin 2\theta.\label{L_dot2}
\end{align}
It is useful to consider in addition the energy of the motion along the y direction, $\epsilon_y=(Y^2+\dot{Y}^2)/2$, which is also a constant of motion of the harmonic oscillator. We find to lowest order

\begin{align}
   \langle \epsilon_y \rangle &=\frac{a^2 \sin^2\theta+b^2\cos^2\theta}{2},\label{epsy}\\
    \langle\dot\epsilon_y \rangle&=  \langle a(1+Y)\dot{X}^2\dot{Y}-X^2\dot{Y}\rangle=-\frac{ab}{8}(a^2-b^2)\sin 2\theta.
\end{align}

and hence we have the identity

\begin{align}
    \epsilon \langle \dot{\epsilon_y}\rangle&= -3L\langle \dot{L}\rangle.
\end{align}

Using the fact that the energy is constant and that $L(\xi=0)=0$, we can integrate this and find
\begin{align}
\epsilon[\langle\epsilon_y(\xi=0)\rangle-\langle \epsilon_y(\xi)\rangle]&= \frac{3}{2}L(\xi)^2.\label{Lsquared}
\end{align}
We have checked numerically that this equation holds for arbitrary initial polar angles as long as $E(0)\ll1$. To obtain a closed-form solution, we additionally assume a small initial angle of the light polarization with the $x$ axis, $\theta_0\ll1$. As one can check {\it a posteriori}, this also implies that $\theta\ll1$ and $|b|\ll |a|$ at all times. In this limit, we can write $\epsilon=a^2/2$ and hence $a(\xi)=a_0$ remains constant. Moreover, we can approximate $\langle\epsilon_y(0)\rangle=a_0^2\theta_0^2/2$ (because $b(0)=0$) and combining Eqs.~(\ref{L_dot2}) and (\ref{epsy}) we obtain
\begin{align}
a_0^2\langle\epsilon_y(\xi)\rangle\simeq \frac{a_0^4 \theta(\xi)^2+a_0^2b(\xi)^2}{2}\simeq \frac{24^2}{2a_0^4}\langle\dot{L}(\xi)\rangle^2+\frac{L(\xi)^2}{2}.
\end{align}
Plugging this relation into Eq.~(\ref{Lsquared}) we arrive at
\begin{align}
 L(\xi)^2+\Bigl(\frac{24\langle\dot{L}(\xi)\rangle}{\sqrt{7}a_0^2}\Bigl)^2=\frac{a_0^4\theta_0^2}{7},
\end{align}
which yields the solution
\begin{align}
    L(\xi)=a_0b(\xi)&=\frac{a_0^2\theta_0}{\sqrt{7}}\sin\omega \xi,\qquad \omega=a_0^2\sqrt{7}/24,\label{Lsol1}\\
  \theta(\xi)&=\theta_0\cos\omega \xi.\label{theta_sol}
\end{align}
These expressions are in quantitative agreement with the numerical results at small angles shown in Fig.~\ref{fig:smallE}. For larger initial angles, there is still qualitative agreement, although the period is somewhat longer and the angular momentum and the polar angle deviate from a single-harmonic time dependence.

\end{document}